\documentclass[12pt]{article}
\input{epsf}  

\title{Pion electromagnetic formfactor in the space-like
region and $P$-phase $\delta^1_1(s)$ of $\pi\pi$ scattering from the
value of the modulus of formfactor in the time-like region.}
\author{B.V.Geshkenbein\thanks{e-mail:geshken@vxitep.itep.ru}\\
Institute for Theoretical and Experimental Physics,\\
su-117279 Moscow Russia}
\date{}
\begin{document}
\maketitle

\newcommand{\be}{\begin{equation}}
\newcommand{\ee}{\end{equation}}

\def\la{\mathrel{\mathpalette\fun <}}
\def\ga{\mathrel{\mathpalette\fun >}}
\def\fun#1#2{\lower3.6pt\vbox{\baselineskip0pt\lineskip.9pt
\ialign{$\mathsurround=0pt#1\hfil##\hfil$\crcr#2\crcr\sim\crcr}}}

\begin{abstract}

The  problem of determining the pion electromagnetic formfactor
$F(q^2)$ in the space-like region from the value of its
modulus in the time-like region is solved by the formfactor analyticity.
 If  $F(q^2)$has no zeroes in the complex plane then the formfactor in
the  space-like region is determined uniquely.  If  $F(q^2)$has zeroes
in the complex plane $q^2$  it can be obtained in the space-like
region within narrow limits using experimental data from the time-like
region.  The formfactor phase $\varphi(s)$ which coincides with the $P$-wave
phase $\delta^1_1(s)$ of the $\pi\pi$ scattering is calculated. The value of
the pion radius has been improved.

\end{abstract}

\newpage

\hspace{2cm}{\bf 1. Introduction.}

\vspace{3mm}
The pion electromagnetic formfactor $F(q^2)$  is calculated theoretically
only in the space-like region. For example, the formfactor has been
calculated with QCD sum rules [1] and in the lattice QCD [2].

On the other hand, the main part of experimental data on  formfactor are
obtained in the time-like region via the reaction
[3,4]

$$e^+e^- \to \pi^+\pi^- \eqno{(a)}$$

At small space-like momentum transfers ($0<Q^2 <0.253~GeV^2$, $Q^2=-q^2$)
the formfactor has been measured by scattering of $300~GeV$ pions from
atomic electrons [5]

$$\pi^+e^- \to \pi^+e^- \eqno{(b)}$$
Colliding--beam measurements of $\sigma(e^+e^- \to \pi^+\pi^-)$ and
measurements of $\sigma(e^+e^- \to \pi^+\pi^-)$ provide direct access to
$F(q^2)$. At large space-like momentum transfers formfactor $F(q^2)$ was
extracted from the reaction of the electroproduction of pions from  nucleons
[6-9]

$$ep \to e\pi^+ n$$

$$en \to e\pi^-p \eqno{(c)}$$
But the presence of the nucleons and their structure complicates theoretical
models and thus the determination of the pion formfactor at large $Q^2$ is
model dependent. Analyticity connects values of the pion formfactor in
space-like and time-like regions.

As follows from microcausality the pion electromagnetic formfactor
$F(q^2)$ has the following analytical properties: 1) $F(q^2)$ is an
analytical function of $q^2$  with a cut along positive $q^2$  from
$q^2=4m^2_{\pi}$ to infinity. 2) On the real axis to the left of
$q^2=4m^2_{\pi}$ the function $F(q^2)$ is real, and consequently, takes
complex conjugate values on the upper and lower edges of the cut. 3)
For complex $q^2$  with $\mid q^2\mid \to \infty$  the function $F(q^2)$
grows no faster than a finite power (in QCD formfactor decreases). 4) the
function $F(q^2)$ is normalized by the condition $F(0)=1$.

It follows from unitarity that the formfactor phase
$\varphi (s)=arctan\frac{Im~F(s)}{Re~F(s)}$ in the region
$4m^2_{\pi}\leq s\leq s_{el}$ must coincide with the $p$-phase of $\pi\pi$
scattering $\delta^1_1(s)$, $s_{el}\approx 0.8~GeV^2$.

The purpose of this work is to calculate formfactor in the space-like region from
the known value of the modulus of the formfactor in the time-like region
measured in [3,4]. The phase of formfactor $\varphi(s)$ which coincides with
$p$-phase $\delta^1_1(s)$  of $\pi\pi$ scattering $\delta^1_1(s)$ measured
in [10-12] will be also calculated.

\bigskip
{\bf 2. Determination of the formfactor $F(s)$ in the whole complex plane
from the given value of its modulus in the time-like region if formfactor
has no zeroes in the complex $s$-plane.}

\vspace{3mm}
We shall follow the method of the works [13]. If $F(s)$ has no zeroes in
complex $s$-plane the function $ln~F(s)$ is an analytical function of $s$
with the cut $[s_0,\infty]$, $s_0=4m^2_{\pi}$ and, consequently, we have the
formula

\be
\frac{1}{2\pi i}\int\limits_C \frac{ln~F(s^{\prime})ds^{\prime}}
{\sqrt{s^{\prime}-s_0}(s^{\prime}-s)s^{\prime}} =
\frac{ln~F(s)}{s\sqrt{s-s_0}}= \frac{1}{2\pi i}\int\limits^{\infty}_{s_0}
\frac{ln\mid F(s^{\prime})\mid^2ds^{\prime}}{\sqrt{s^{\prime}-s_0} s^{\prime}
(s^{\prime}-s)}
\ee
where the contour $C$ contains the lower and upper edges of the cut and a
large circle. It follows from Eq.(1) that when $F(s)$ has no zeroes in the
complex $s$-plane the function $F(s)$ is uniquely determined

$$ F(s) = F_0(s)$$

\be
F_0(s) = exp \Biggl [ \frac{s\sqrt{s_0-s}}{2\pi}\int\limits^{\infty}_{s_0}
\frac{ln\mid F(s^{\prime})\mid^2 ds^{\prime}}{\sqrt{s^{\prime}-s_0}
s^{\prime}(s^{\prime} -s)}\Biggr ]
\ee
The result of the calculation of $F_0(s)$ is given in tables 1,2 and in
Fig.1. We use for \\
$\mid F(s^{\prime}) \mid ~~ s_0 < s^{\prime} <
\infty$ the phenomenological formula for formfactor obtained in [14,15]. This
formula satisfies the following requirements: (1) formfactor has correct
analytical properties; (2) at large $Q^2 = -s$ formfactor has the asymptotic
behaviour determined by QCD [16-19] (taking into account the preasymptotic
power correction [20]);  the phenomenological formula  describes well
the experiments [3,4] $\chi = 123$ fitting over 120 experimental points in
the interval $0.1296~GeV^2 \leq s\leq 4.951~GeV^2$.

The phase of the formfactor can be found using formula

\be
\frac{1}{s^{\prime}-s-i\varepsilon}=\pi i~\delta(s^{\prime}-s) + P\frac{1}
{s^{\prime}-s}
\ee
Substituting Eq.(3)  into Eq.(2) we get

\be
F_0(s) = \mid F(s)\mid e^{i\varphi_0 (s)} , ~~~~~~ s > s_0
\ee
where the phase of the formfactor $\varphi_0(s)$ is equal to

\be
\varphi_0(s)   = -\frac{s\sqrt{s-s_0}}{2\pi}P\int\limits^{\infty}_{s_0}
\frac{ln\mid F(s^{\prime})\mid^2 ds^{\prime}}{\sqrt{s^{\prime}-s_0}s^{\prime}
(s^{\prime}-s)}, ~~~~~~s > s_0
\ee
Let us write the integral (5) in the form more convenient for numerical
calculation

\be
\varphi_0(s)=-\frac{s\sqrt{s-s_0}}{2\pi}\left \{
\int\limits^{\infty}_{s_0} \frac{ln\mid
F(s^{\prime})/F(s)\mid^2ds^{\prime}}{\sqrt{s^{\prime}-s_0}s^{\prime}
(s^{\prime}-s)}- \frac{\pi}{s\sqrt{s_0}}ln\mid F(s)\mid^2 \right \}
\ee
The result of the calculations of $\varphi_0(s)$ is given in the Table 1.

\bigskip

{\bf 3. The consideration of the formfactor with zeroes in the complex plane.}

\vspace{3mm}
Let us consider now the case when the formfactor $F(s)$ has zeroes in the
complex plane at $s=s_k$, $k=1,2,... .$ We introduce instead of the
formfactor $F(s)$ the function $\tilde{F}(s)$ by formula

$$F(s) =\chi(s)\tilde{F} (s)$$

\be
\chi(s) = \prod_k~\chi_k(s)
\ee

$$\chi_k(s) = \frac{\sqrt{s_0-s}-\sqrt{s_0-s_k}}{\sqrt{s_0 -s}+
\sqrt{s_0-s_k}}\cdot \frac{\sqrt{s_0-s}-\sqrt{s_0-s_k^*}}
{\sqrt{s_0-s}+\sqrt{s_0-s^*_k}}$$
Since to each complex zero $s_k$ there corresponds a complex  conjugate zero
$s^*_k$, the modulus of the function $\chi(s)$ equals  to unity on the cut.
The function $\tilde{F}(s)$  has no zeroes in the complex plane, on the cut
its modulus equals to $\mid F(s^{\prime})\mid$, and it is normalized by the
condition

\be
\tilde{F}(0) = \chi^{-1}(0)\geq 1
\ee
and, consequently, we may apply the Cauchy theorem to the function
$[ln~\tilde{F}(s)]/s\cdot \sqrt{s-s_0}$

$$\frac{1}{2\pi i}\int\limits_C \frac{ln~\tilde{F}(s^{\prime})ds^{\prime}}
{\sqrt{s^{\prime}-s_0}(s^{\prime}-s)s^{\prime}} =
\frac{ln~\tilde{F}(s)}{s\sqrt{s-s_0}} +
\frac{ln~\tilde{F}(0)}{i(-s)\sqrt{s_0}} =$$

\be
=  \frac{1}{2\pi i}\int\limits_C^{\infty}
\frac{ln~\mid F(s^{\prime})\mid^2 ds^{\prime}}
{\sqrt{s^{\prime}-s_0} s^{\prime}(s^{\prime}-s)}
\ee
Therefore

$$ F(s) =\psi(s) F_0(s)$$

\be
\psi(s) = \chi(s)/[\chi (0)]^{\sqrt{1-s/s_0}}
\ee
where $F_0(s)$ is determined by Eq.(2).

It is clear from Eq.(7) that the knowledge of the modulus of the
formfactor on the
cut determines the value of the formfactor in the entire complex plane up to
within the factor $\psi(s)$.

It is easy to obtain from Eq.(10) the asymptotic formula for the formfactor
$F(s)$
for $s\to -\infty$. Since $\chi(-s)\to 1$  as $s\to \infty$, we have

\be
\mid F(-s)\mid  \to \mid F(s)\mid exp(-[a +ln~\chi(0)] \cdot \sqrt{s/s_0})
\ee
where

\be
a= \frac{\sqrt{s_0}}{2\pi} \int\limits^{\infty}_{s_0} \frac{ln~\mid
F(s)\mid^2 ds}{\sqrt{s-s_0}~s}
\ee
From QCD there follows the asymptotic formula
at $Q^2=-q^2 \to \infty$ for the pion formfactor [16-19]

\be
F(-Q^2) \to \frac{16\pi \alpha_s (Q^2)}{Q^2}f^2_{\pi}
\ee
$f_{\pi}=93~MeV$. Therefore

\be
a + ln~\chi(0) =0
\ee
If the formfactor $F(s)$ has no zeroes in the complex  plane than $a=0$  and
the formfactor $F(s)$ can be determined uniquely. If the value $a$  is small
than the uncertainty of the determination of the formfactor $F(s)$ due to
the function $\psi(s)$  will be small. The value $a$  was calculated in the
work [15] from the analysis of the experiments [3,4] and was found to be
small:

\be
a = 0.069 \pm 0.003
\ee

\bigskip
{\bf 4. Bounds on the formfactor $F(s)$ in the space-like region.}

\vspace{3mm}
The formfactor $F(s), s< 0$ is determined up to  the factor $\psi(s)$. To
determine upper and lower bounds on the value of the formfactor 
$F(s)$, for $s < 0$  we look for the maximum and the minimum 
of the factor $\psi(s)$, by
fixing the value of $s$  and changing the distribution of the zeroes $s_k$
so that condition (14)  is satisfied. Let us write the relation

\be
w_k = (\sqrt{s_0} - \sqrt{s_0-s_k})/(\sqrt{s_0} + \sqrt{s_0 - s_k})
\ee
Write $w_k$  in the form:

\be
w_k = r_k e^{i\varphi_k}
\ee
$r_k<1$.

Let us write the factor $\chi_k(s)$  in Eq.(7) in the form

\be
\chi_k(s) = \frac{(w-w_k)(w-w^*_k)}{(1-ww_k)(1-ww^*_k)} = \frac{w^2-2w r_k
cos\varphi_k + r^2_k}{1- 2wr_k cos\varphi_k + w^2 r^2_k}
\ee
where

\be
w = (\sqrt{s_0}-\sqrt{s_0 -s})/(\sqrt{s_0}+\sqrt{s_0-s})
\ee
It is obvious that the maximum $\chi_k(s)$ is achieved if $cos \varphi_k=1$
and the minimum $\chi_k(s)$  is achieved if $cos\varphi_k=-1$  and

\be
max ~\chi_k(s) = \frac{(w-r_k)^2}{(1-w r_k)^2}
\ee

\be
min ~\chi_k(s) = \frac{(w+r_k)^2}{(1+w r_k)^2}
\ee
The value $\chi(0)$ is equal

\be
\chi(0) = \prod_k r^2_k
\ee
It follows from Eq.(14)

\be
\chi(0) = e^{-a}
\ee
And it follows from inequality

\be
\frac{\mid w \mid +r_1}{1+\mid w \mid r_1}~\frac{\mid w \mid+r_2}{1+\mid w
\mid r_2}< \frac{\mid w \mid + r_1r_2}{1+\mid w \mid r_1 r_2}
\ee
and from Eqs. (22,23)  that the maximum $\psi(s)$ is achieved if $F(s)$
has one double real zero

$$w_{max}=e^{-a/2},
~~~s_{max}=\frac{4e^{-a/2}}{(1+e^{-a/2})^2}s_0=0.9997~s_0 $$
Inequalities $r_1 > \mid w \mid,~ r_2 > \mid w \mid,~ r_1, r_2 > \mid w
\mid$

\be
\frac{r_1 -\mid w \mid}{1-\mid w \mid r}~\frac{r_2 -\mid  w
\mid}{1-\mid w \mid r} > \frac{r_1r_2 -\mid w \mid}{1-\mid w \mid r_1 r_2}
\ee
and  Esq. (22,23) prove that minimum $\psi(s)$  is achieved if $F(s)$  has
one double  real zero

$$w_{min}=-e^{-a/2},~~~~s_{min}=-\frac{4 e^{-a/2}}{(1-e^{-a/2})^2}s_0=
-3360.31~s_0\footnote{Strictly speaking,  minimum and maximum
$\psi(s)$ are achieved at simple real zero. But it practically don't change
all results.}$$

\bigskip
{\bf 5. The contribution of complex zeroes into the phase of
the pion formfactor $\varphi(s)$.}

\vspace{3mm}
The contribution of complex zeroes in the phase of the formfactor
$\varphi(s)$ is defined by the formula

\be
\delta\varphi(s) = \frac{1}{2 i}ln~\psi(s), ~~~ s > s_0
\ee
The contribution of pair complex conjugate zeroes in phase of the function
$\chi_k$  can be obtained from Eq.(18)  putting $w=(1-0)e^{i\theta}$

\be
\chi_k(s) =e^{2i(\theta+\theta_k)}
\ee
where

\be
\theta = -i~ln\frac{1+i\sqrt{s/s_0-1}}{1-i\sqrt{s/s_0-1}}=2
arc~sin\sqrt{1-s_0/s}
\ee

\be
\theta_k = \frac{1}{2i}ln\frac{1-2 r_k e^{-i\theta}cos \varphi_k + r^2_k
e^{-2i\theta}}{1 - 2r_k e^{i\theta} cos \varphi_k + r^2_k e^{2i\theta}}
\ee
Maximum and minimum of the function $\theta_k$ by $\varphi_k$ are achieved
if cos $\varphi_k=\pm 1$.

\be
\theta^{\pm}_k = \frac{1}{i}ln \frac{1\mp r_k e^{-i\theta}}
{1 \mp r_k e^{i\theta}}
 = \pm 2 arc~sin \frac{r_k sin~ \theta}{\sqrt{1\mp 2 r_k cos
\theta + r^2_k}}
\ee
The contribution of pair complex conjugate the zeroes satisfying the
condition (23) has the form

\be
\delta \varphi^{(\pm)} (s) = 2(\theta + \theta^{(\pm)}_2 + \delta_0)
\ee
where $\theta_2$ follows from Eq.(30) by changing $r_k \to r =e^{-a/2}$ and
$\delta_0 = -\frac{a}{2}\sqrt{s/s_0 -1}$.

In Table 2 we give the values of $\varphi_0$, $\delta\varphi^{(+)}$,
$\delta\varphi^{(-)}$ and the values of the $P$-phase $\pi\pi$-scattering
$\delta^1_1(s)$ from [11]. It is seen from Table 3 that there is a very good
lower bound on the phase $\varphi(s)$. Upper bound on the phase $\varphi(s)$
is practically absent.

\bigskip
{\bf 6. Improved determination of upper bound of the
formfactor in the space-like
region.}

\vspace{3mm}
It can be seen from Table 3 that $\varphi_0(s)$  within the limits of
experimental errors coincides with $\delta^1_1(s)$. This means that the value
$\delta \varphi^+(s)$ has the order of the experimental error in
$\delta^1_1(s)$.
Thus we change $\theta^{(+)}_k$ on $\theta_k$  and $cos \varphi_k=1$  on
$cos\varphi_k=-0.96$. If $-1\leq cos \varphi_k \leq -0.96$  the phase
$\varphi(s)$ coincides with $\delta^1_1(s)$  to within the experimental
errors. Improved upper bound of the formfactor in the space-like region is obtained from
Eq.(18) if $(cos\varphi_k)_{max.impr.}=-0.96$.

The results of the calculation of $F_0,F_{min}, F_{max}, F_{max.,impr.}$ and
the experimental data from ref.[5-9]  are shown in Tables 1,2 and Fig.
The curves $F_0(s)$ and $F_{min}(s)$ merge together.

\vspace{5mm}
{\bf 7. Improved calculation of the pion radius.}

\vspace{3mm}
The formulae for the bounds on the pion radius were obtained in [13,21,22]

\be
(r^2_{\pi})_{max} = \frac{3}{2m^2_{\pi}} \Biggl [b + \frac{1}{2} (sha -
a) \Biggr ]
\ee

\be
(r^2_{\pi})_{min} = \frac{3}{2m^2_{\pi}} \Biggl [b - \frac{1}{2} (sha + a)
\Biggr ]
\ee
where

\be
b = \frac{s^{3/2}_0}{2 \pi} \int \limits ^{\infty}_{s_0}~ \frac{ln \mid
F(s) \mid ^2 ds}{s^2 \sqrt{s - s_0}}
\ee
The value $b$ was calculated in [15]

\be
b = 0.1544 \pm 0.0016
\ee
Formulae (32, 33) were obtained   from the derivative of the formfactor
$F(s)$  with respect to $s$ at $s=0$

\be
F^{\prime}(0) = \Biggl ( b - \sum \limits_{k}~\frac{1 - w_k}{4 w_k} +
\frac{1}{2} ln \mid \prod_k w_k \mid\Biggr ) /s_0
\ee
and by definition

\be
r^2_{\pi} = 6 \cdot  F^{\prime}(0)
\ee
The maximum value of $F^{\prime}(0)$ is reached when the function
$F(s)$ has one negative zero $(s_1)_{max}$, so that $(w_1)_{max} =
-e^{-a}$ and then $(s_1)_{max} = -839.8 s_0$, $(r^2_{\pi})_{max} =
(0.463 \pm 0.005) fm^2$. The minimum value of $F^{\prime}(0)$
is reached when the function $F(s)$ has one positive zero
$(s_1)_{min}$, so that $(w_1)_{min} = e^{-a}$ and then $(s_1)_{min} = 0.9988
s_0$, $(r^2_{\pi})_{min} = 0.256 fm^2$.  This zero gives in the phase
$\varphi (s)$ the additional term $\sim 180^0$ what is inconsistent with the
experimental data on $\delta^1_1(s)$ [11]. The minimum of $F^{\prime}(0)$
which is consistent with the experimental data of $\delta^1_1(s)$
 is reached
when  formfactor $F(s)$ has two complex conjugate zeroes $(w_1)_{min.impr} =
r_1 e^{i \varphi}$,~~ $(w^*_1)_{min.impr.} = r_1 e^{-i \varphi_1}$, and
$(r_1)_{min.impr.} = e^{-a/2}$,~~ $(cos \varphi_1)_{min.impr.} = -0.96$,~
$(s_1)_{min.impr.} = (46.23 + 11.33 i)s_0$.

We have obtained:

\be
F^{\prime}_{min.impr.} (0) = b - \frac{a}{2} + 0.96 sh \frac{a}{2} =
0.1530 \pm 0.0016
\ee
and

\be
(r^2_{\pi})_{min.impr.} = (0.4623 \pm 0.0048) fm^2
\ee
Taking into account the closeness of $(r^2_{\pi})_{max}$ and
$(r^2_{\pi})_{min.impr.}$ we have obtained:

\be
r^2_{\pi} = (0.463 \pm 0.005) fm^2
\ee
This value of $r^2_{\pi}$ is slightly larger than those obtained in [3,
5]

$$
r^2_{\pi} = (0.422 \pm 0.003 \pm 0.013) fm^2 ~~~~~~~~~~~~~~~~~~[3]
$$
\be
~~~~~~~~~ = (0.439 \pm 0.008) fm^2 ~~~~~~~~~~~~~~~~~~~~~~~[5]
\ee
This disagreement is due to the fact that the authors of [3,5] used models,
which give the underestimated value of $r^2_{\pi}$ [15].

I thank V.L.Morgunov and V.A.Novikov for useful discussions.

\newpage

\hspace{10.5cm}  {\bf Table 1}

\vspace{2mm}
\begin{center}
The result of the calculations of the pion formfactor
$F(-Q^2)$ in the space-like region $0\leq Q^2 \leq 0.253 ~GeV^2$
\end{center}

\vspace{3mm}
\begin{tabular}{|c|c|c|c|c|c|c|} \hline
$n$ & $Q^2$ & $F^2(-Q^2)_{Exp}$ & $F^2_0(-Q^2)$ & $F^2_{min}(-Q^2)$ &
$F^2_{max}(-Q^2)$ & $F^2_{ max,impr.}(-Q^2)$ \\ \hline
0 & 0 & 1 & 1 & 1 & 1 & 1 \\ \hline
1 & 0.015 & $0.944\pm0.007$  & 0.943 & 0.943 & 0.966 & 0.944 \\ \hline
2 & 0.017 & $0.921 \pm 0.006$ & 0.935 & 0.935 & 0.961 & 0.936 \\ \hline
3 & 0.019 & $0.933 \pm 0.006$ & 0.928 & 0.928 & 0.957 & 0.929 \\ \hline
4 & 0.021 & $0.926 \pm 0.006$ & 0.921 & 0.921 & 0.952 & 0.922 \\ \hline
5 & 0.023 & $0.914 \pm 0.007$ & 0.914 & 0.914 & 0.948 & 0.915 \\ \hline
6 & 0.025 & $0.905 \pm 0.007$ & 0.907 & 0.907 & 0.943 & 0.908 \\ \hline
7 & 0.027 & $0.898 \pm 0.008$ & 0.900 & 0.900 & 0.938 & 0.901 \\ \hline
8 & 0.029 & $0.884 \pm 0.008$ & 0.894 & 0.894 & 0.934 & 0.895 \\ \hline
9 & 0.031 & $0.884 \pm 0.009$ & 0.887 & 0.887 & 0.929 & 0.888 \\ \hline
10 & 0.033 & $0.890 \pm 0.009$ & 0.881 & 0.881 & 0.925 & 0.882 \\ \hline
11 & 0.035 & $0.866 \pm 0.010$ & 0.871 & 0.871 & 0.917 & 0.872 \\ \hline
12 & 0.037 & $0.876 \pm 0.011$ & 0.868 & 0.868 & 0.916 & 0.869 \\ \hline
13 & 0.039 & $0.857 \pm 0.011$ & 0.861 & 0.861 & 0.911 & 0.862 \\ \hline
14 & 0.042 & $0.849 \pm 0.009$ & 0.852 & 0.852 & 0.905 & 0.854 \\ \hline
15 & 0.046 & $0.837 \pm 0.009$ & 0.840 & 0.840 & 0.896 & 0.842 \\ \hline
16 & 0.050 & $0.830 \pm 0.010$ & 0.828 & 0.828 & 0.887 & 0.830 \\ \hline
17 & 0.054 & $0.801 \pm 0.011$ & 0.816 & 0.816 & 0.878 & 0.818 \\ \hline
18 & 0.058 & $0.800 \pm 0.012$ & 0.805 & 0.870 & 0.870 & 0.807 \\ \hline
19 & 0.062 & $0.809 \pm 0.012$ & 0.793 & 0.793 & 0.861 & 0.795 \\ \hline
20 & 0.066 & $0.785 \pm 0.014$ & 0.782 & 0.782 & 0.853 & 0.784 \\ \hline
21 & 0.070 & $0.785\pm 0.015 $ & 0.772 & 0.772 & 0.844 & 0.775 \\ \hline
22 & 0.074 & $0.777 \pm 0.016$ & 0.761 & 0.761 & 0.836 & 0.764 \\ \hline
23 & 0.078 & $0.769 \pm 0.017$ & 0.751 & 0.751 & 0.828 & 0.754 \\ \hline
24 & 0.083 & $0.757 \pm 0.010$  & 0.738 & 0.738 & 0.818 & 0.741  \\ \hline
25 & 0.089 & $0.715 \pm 0.016$ & 0.724 & 0.724 & 0.806 & 0.727 \\ \hline
26 & 0.095 & $0.724 \pm 0.018$ & 0.710 & 0.710 & 0.795 & 0.714 \\ \hline
27 & 0.101 & $0.680 \pm 0.017$ & 0.696 & 0.696 & 0.783 & 0.700 \\ \hline
28 & 0.107 & $ 0.696 \pm 0.019$ & 0.683 & 0.683 & 0.772 & 0.687 \\ \hline
29 & 0.013 & $0.688 \pm 0.020$ & 0.670 & 0.670 & 0.761 &  0.674 \\ \hline
30 & 0.119 & $0.676 \pm 0.021$ & 0.657 & 0.657 & 0.750 & 0.661 \\ \hline
31 & 0.125 & $0.665 \pm 0.023 $ & 0.645 & 0.645 & 0.740 & 0.650 \\ \hline
32 & 0.131 & $0.651 \pm 0.024$ & 0.633 & 0.633 & 0.729 & 0.638 \\ \hline
33 & 0.137 & $0.646 \pm 0.027 $& 0.621 & 0.621 & 0.719 & 0.626 \\ \hline
34 & 0.144 & $0.616 \pm 0.023$ & 0.608 & 0.608 & 0.708 & 0.613 \\ \hline
35 & 0.153 & $0.654\pm 0.023$ & 0.592 & 0.592 & 0.693 & 0.597 \\ \hline

\end{tabular}

\newpage

\begin{tabular}{|c|c|c|c|c|c|c|} \hline
$n$ & $Q^2$ & $F^2(-Q^2)_{Exp}$ & $F^2_0(-Q^2)$ & $F^2_{min}(-Q^2)$ &
$F^2_{max}(-Q^2)$ & $F^2_{max,impr.}(-Q^2)$ \\ \hline
36 & 0.163 & $0.563 \pm 0.024$ & 0.575 & 0.575 & 0.677 & 0.581 \\ \hline
37 & 0.173 & $0.534 \pm 0.038$ & 0.558 & 0.558 & 0.662 & 0.564 \\ \hline
38 & 0.183 & $0.586 \pm 0.034$ & 0.542 & 0.542 & 0.648 & 0.548 \\ \hline
39 & 0.193 & $0.544 \pm 0.036$ & 0.527 & 0.527 & 0.634 & 0.533 \\ \hline
40 & 0.203 & $0.529 \pm 0.040$ & 0.513 & 0.513 & 0.620 & 0.520 \\ \hline
41  & 0.213 & $0.616 \pm 0.048$ & 0.499 & 0.499 & 0.607 & 0.506 \\ \hline
42 & 0.223 & $0.487 \pm 0.049$ & 0.486 & 0.485 & 0.594 & 0.493 \\ \hline
43 & 0.233 & $0.417 \pm 0.058$ & 0.473 & 0.473 & 0.581 & 0.480 \\ \hline
44 & 0.243 & $0.593 \pm 0.074$ & 0.460 & 0.460 & 0.569 & 0.468 \\ \hline
45 & 0.253 & $0.336 \pm 0.073$ & 0.449 & 0.448 & 0.558 & 0.457 \\ \hline
\end{tabular}

\bigskip
\noindent
1. $F_0(-Q^2)$ is free from complex zeroes formfactor.\\
2. The formfactor $F_{min}(-Q^2)$  has complex zeroes so that $F(-Q^2)$ is
minimal.\\
3. The formfactor $F_{max}(-Q^2)$  has complex zeroes so that $F(-Q^2)$ is
maximal.\\
4. The formfactor $F_{max.impr}(-Q^2)$ has complex zeroes so $F(-Q^2)$ is
maximal and the phase of the formfactor $\varphi(s)$
coincides with $p$-phase $\delta^1_1(s)$ of the $\pi\pi$-scattering
ref.[11].\\
5. $F_{exp}(-Q^2)$ is the experimental value of the formfactor  from ref [5].
$F_{exp}(Q^2)$ is measured by scattering  300 GeV pions
from the electrons of a liquid hydrogen target.

\newpage
\hspace{10cm}  {\bf Table 2}

\vspace{2mm}

\begin{center}
The results of the calculations of the pion formfactor $F(-Q^2)$
in the space-like region $0.18 \leq Q^2 \leq 9.77~GeV^2$.
\end{center}

\vspace{5mm}

\begin{tabular}{|l|l|l|l|l|c|} \hline
\multicolumn{6}{|c|}{Ref. [6]}   \\ \hline
 $Q^2/GeV^2$ & ~~$(F(-Q^2))_{Exp}$ & ~~$F_0(-Q^2)$ & ~~$F_{min}(-Q^2)$ &
~~$F_{max}(-Q^2)$ & $F_{max.imp~r}(-Q^2)$ \\ \hline
0.18 & $0.850 \pm 0.044$ & 0.740 & 0.740 & 0.807 & 0.744\\
\hline
0.29 & $0.634 \pm 0.029$ & 0.639 &
0.639 & 0.719 & 0.646 \\  \hline
0.40 & $0.570\pm 0.016$ & 0.562 & 0.562 &
0.649 & 0.571 \\ \hline
0.79 & $0.384 \pm 0.014$ & 0.391 & 0.391 & 0.483 &
0.407 \\ \hline
1.19 & $0.238 \pm 0.017$ & 0.295 & 0.295 & 0.383 & 0.315 \\
\hline
\multicolumn{6}{c}{Ref. [7]}   \\ \hline
0.62 & $0.445 \pm 0.016$ &
0.452 & 0.452 & 0.543 & 0.465 \\ \hline
1.07 & $0.309 \pm 0.019$ & 0.319 &
0.319 & 0.409 & 0.338 \\ \hline
1.20 & $0.269 \pm 0.012$ & 0.293 & 0.293 &
0.381 & 0.313 \\ \hline
1.31 & $0.242\pm 0.015$ & 0.274 & 0.274 & 0.361 &
0.295 \\  \hline
1.20 & $0.262 \pm 0.014$ & 0.293 & 0.293 & 0.381 & 0.313 \\
\hline
2.01 & $0.154 \pm 0.014$ & 0.191 & 0.191 &0.270 & 0.216 \\ \hline
\multicolumn{6}{c}{Ref.[8]}   \\ \hline
1.22 & $0.290 \pm 0.030$ & 0.290 & 0.289 & 0.378 & 0.310 \\ \hline
1.20 & $0.294 \pm 0.019$ & 0.293 & 0.293 & 0.381 & 0.313 \\ \hline
1.71 & $0.238 \pm 0.020$ & 0.221 & 0.229 & 0.303 & 0.245 \\ \hline
3.30 & $0.102 \pm 0.023$ & 0.118 & 0.117 & 0.184 & 0.146 \\ \hline
1.99 & $0.179 \pm 0.021$ & 0.193 & 0.193 & 0.272 & 0.218 \\ \hline
3.99 & $0.004\pm 0.678$ & 0.096 & 0.095 & 0.157 & 0.124 \\ \hline
\multicolumn{6}{c}{Ref. [9]}   \\ \hline
1.18 & $0.256\pm 0.026$ & 0.297 & 0.297 & 0.385 & 0.317 \\ \hline
1.94 & $0.193 \pm 0.025$ & 0.198 & 0.197 & 0.277 & 0.223 \\ \hline
3.33 & $0.086 \pm 0.033$ & 0.117 & 0.116 & 0.183 & 0.145 \\ \hline
6.30 & $0.059 \pm 0.030$ & 0.056 & 0.055 & 0.105 & 0.083 \\ \hline
9.77 & $0.070 \pm 0.019$ & 0.032 & 0.030 & 0.068 & 0.056 \\ \hline

 \end{tabular}


\vspace{3mm}

\noindent
1.$F_0(-Q^2)$  is free from complex zeroes formfactor.\\
2. The formfactor $F_{min}(-Q^2)$ has complex zeroes so that $F(-Q^2)$ is
minimal.\\
3. The formfactor $F_{max}(-Q^2)$  has complex zeroes so that $F(-Q^2)$ is
maximal.\\
4. The formfactor $F_{max.impr.} (-Q^2)$ has complex zeroes, so $F(-Q^2)$ is
maximal and the phase of $\varphi(s)$ coincides with $p$-phase
$\delta^1_1(s)$ of the $\pi\pi$-scattering [ref.11].\\
5. $F_{exp}(-Q^2)$ is the experimental value of the formfactor from
ref.[6-9].  $F(-Q^2)_{Exp}$ has been obtained from the electroproduction of
pions from nucleons $ep \to e \pi^+ n$ by extrapolation.  

\newpage
\hspace{9cm}  {\bf Table 3}

\vspace{2mm}

\begin{center}
The results of the calculations of the phase $\varphi(s)$
of the pion formfactor.
\end{center}

\vspace{5mm}

\begin{tabular}{|c|l|l|c|c|} \hline
$\sqrt{s}/GeV$ & $\varphi_0(s)$/deg. & $\delta\varphi^{(+)}(s)/deg.$ &
$\delta\varphi^{(-)}(s)/deg.$ & $\delta^1_1(s)/deg.$\\  \hline
0.51 & 9.86 & 351.35 & -0.0020 & $9.3\pm 0.7 $ \\ \hline
0.53 & 11.31 & 351.15 & -0.0023 & $10.4 \pm 0.6$ \\ \hline
0.55 & 12.95 & 350.94 & -0.0026 & $13.1 \pm 0.8$ \\ \hline
0.57 & 14.83 & 350.72 & -0.0029 & $13.5\pm 0.7$ \\ \hline
0.59 & 17.02 & 350.49 & -0.0033 & $17.6 \pm 0.8$ \\ \hline
0.61 & 19.64 & 350.26 & -0.0037 & $19.4 \pm 0.8$ \\ \hline
0.63 & 22.84 & 350.02 & -0.0041 & $20.9\pm 0.8$ \\ \hline
0.65 & 26.70 & 349.78 & -0.0045 & $25.5 \pm 0.7$ \\ \hline
0.67 & 31.71 & 349.53 & -0.0050 & $32.1 \pm 0.7$ \\ \hline
0.69 & 38.31 & 349.28 & -0.0055 & $37.5 \pm 0.5$ \\ \hline
0.71 & 47.07 & 349.03 & -0.0060 & $46.1 \pm 0.9$ \\ \hline
0.73 & 58.65 & 348.52 & -0.0071 & $73.0 \pm 2.3$ \\ \hline
0.75 & 73.28 & 348.52 & -0.0071 & $73.0 \pm 2.3$ \\ \hline
0.79 & 106.12 & 348 & -0.0084 & $113.3 \pm 1.9$ \\ \hline
0.81 & 117.09 & 347.74 & -0.0091 & $118.1 \pm 1.1$ \\ \hline

\end{tabular}

\vspace{3mm}

\noindent
1. $\varphi_0(s)$  is phase of the pion formfactor  free from complex
zeroes.\\
2. $\delta\varphi^+(s)$ is the maximal contribution of complex
zeroes in the phase of the pion formfactor.\\
3. $\delta\varphi^-(s)$  is the
minimal contribution of complex zeroes in the phase of the pion
formfactor.\\
4.  $\delta^1_1$(s) is the $P$-phase of the
$\pi\pi$-scattering [ref.11].

\newpage

\begin{figure}
\centerline{\epsffile{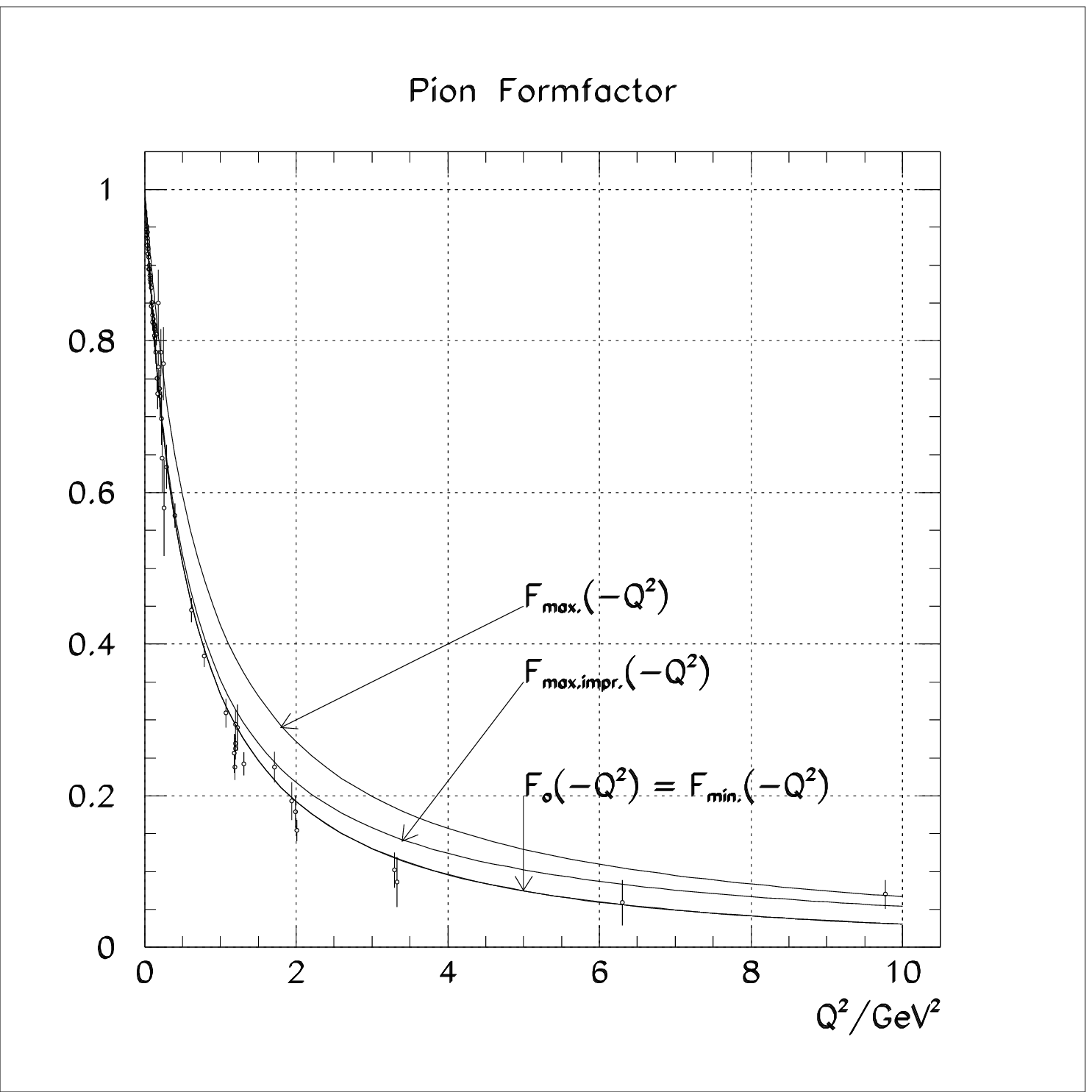} }
\caption{
1. $F_0(-Q^2)$ is free from complex zeroes formfactor.
2. The formfactor $F_{min}(-Q^2)$  has complex zeroes so that $F(-Q^2)$ is
minimal.
3. The formfactor $F_{max}(-Q^2)$  has complex zeroes so that $F(-Q^2)$ is
maximal.
4. The formfactor $F_{max.impr}(-Q^2)$ has complex zeroes so $F(-Q^2)$ is
maximal and the phase of the formfactor $\varphi(s)$
coincides with $p$-phase $\delta^1_1(s)$ of the $\pi\pi$-scattering
ref.[11].
5. The experimental value of the formfactor  from ref [5-9].
}
\label{fig:detect}
\end{figure}

\end{document}